\documentclass[a4paper,twocolumn]{esapub} 
\usepackage{natbib} 
 
\input psfig.sty 
\title{THE INTEGRAL/IBIS TELESCOPE MODELING} 
 
\author{P. Laurent} 
\author{ O. Limousin} 
\affil{DSM/DAPNIA/SAp, CEA/Saclay, France} 
\author{G. Malaguti} 
\author{E. Caroli} 
\affil{ITESRE, Bologne, Italy} 
\author{G. de Cesare} 
\affil{IAS, Rome, Italy} 
\author{A.J. Bird} 
\affil{SOTON, Southampton, UK} 
\author{\\J. Grygorczuk} 
\affil{CBK, Warsaw, Poland} 
\author{J.M. Torrejon} 
\affil{Universitat de Alicante, Alicante, Spain}

\begin{document} 
 
\maketitle 
 
 
\begin{abstract} 
 
The main objective of the IBIS modeling activities is the 
generation of the IBIS spectral response matrix. This generation 
will be done step by step by firstly constructing  ``calibrated 
model" of each IBIS sub-systems. A  ``calibrated model" is defined 
as a GEANT Monte-Carlo model, further checked by intensive 
corresponding calibration. These calibrated models will be in a 
second step integrated in the whole IBIS Mass Model. This Mass 
Model will be checked and refined using the data obtained during 
the IBIS Telescope calibration, the PLGC during the ETET period, 
and the In-flight calibration data acquired during the whole 
INTEGRAL mission. The IBIS Mass Model will be used also for 
computations of efficiencies and transparencies. Some imaging 
studies (such as the study of the diffuse emission) may be also 
done with the help of the IBIS Mass Model. The IBIS Mass Model may 
be used for the simulation of IBIS data in order to check the 
different IBIS software. 
 
\end{abstract} 
 
\section{The INTEGRAL satellite and its instruments}

\noindent INTEGRAL is a 15 keV-10 MeV $\gamma$-ray mission with concurrent 
source monitoring at X-rays (3-35 keV) and in the optical range (V, 500- 
600 nm). All instruments are coaligned and have a large FOV, covering 
simultaneously a very broad range of sources. The INTEGRAL payload consists 
of two main $\gamma$-ray instruments, the spectrometer SPI and the imager IBIS, 
and of two monitor instruments, the X-ray monitor JEM-X and the Optical 
Monitoring Camera OMC. 
 
 
\noindent The Imager on Board Integral Satellite (IBIS) provides diagnostic capabilities 
of fine imaging (12' FWHM), source identification and spectral sensitivity 
to both continuum and broad lines over a broad (15~keV--10~MeV) energy range. 
It has a continuum sensitivity of 2~10$^{-7}$~ph~cm$^{-2}$~s$^{-1}$ at 1~MeV 
for a 10$^6$ seconds observation and a spectral resolution better than 7~$\%$ 
@ 100~keV and of 6~$\%$ @ 1~MeV. The imaging capabilities of the IBIS are 
characterized by the coupling of its source discrimination capability 
(angular resolution 12' FWHM) with a field of view (FOV) of 9$^\circ$ 
$ \times $ 9$^\circ$ fully coded, 29$^\circ$ $ \times $ 29$^\circ$ partially coded FOV.

 
\noindent The spectrometer SPI will perform spectral analysis of $\gamma$ 
ray point sources and extended regions with an unprecedented energy 
resolution of $\sim$ 2 keV (FWHM) at 1.3 MeV. Its large field of view 
(16$^{\circ}$ circular) and limited angular resolution ( 2$^{\circ}$ FWHM) 
are best suited for diffuse sources imaging but it retains nonetheless the 
capability of imaging point sources. It has a continuum sensitivity of 
7 $\times$ 10$^{-8}$ ph cm$^{-2}$ s$^{-1}$ at 1 MeV and a line sensitivity 
of 5$\times$ 10$^{-6}$ ph cm$^{-2}$ s$^{-1}$ at 1 MeV, both 3$\sigma$ for 
a 10$^6$ seconds observation.

 
The Joint European Monitor JEM-X supplements the main INTEGRAL instruments 
and provides images with 3' angular resolution in a 4.8$^{\circ}$ fully coded 
FOV in the 3-35 keV energy band. The Optical Monitoring Camera (OMC) will 
observe the prime targets of INTEGRAL main $\gamma$ ray instruments. Its 
limiting magnitude is M$_V$ $\sim$ 19.7 (3$\sigma$, 10$^3$ s). The wide band observing opportunity offered by INTEGRAL provide for the first 
time the opportunity of simultaneous observing over 7 orders of magnitude. 
 
\section{IBIS modeling} 
 
The generation of the IBIS response matrices will be done step by 
step by firstly constructing ``calibrated model" of each IBIS 
sub-systems. A  ``calibrated model" is defined as a GEANT 
Monte-Carlo model (Brun et al. 1994), further checked by intensive 
corresponding calibration. These calibrated models will be in a 
second step integrated in the whole IBIS Mass Model. This Mass 
Model will be checked and refined using the data obtained during 
the IBIS Telescope calibration, the PLGC during the ETET period 
(Carli et al., 2000), and the In-flight calibration data acquired 
during the whole INTEGRAL mission. We will detail below the status 
of this activity. 
 
\subsection{IBIS QM modeling} 
 
In a first step, we have made a model of the IBIS QM by assembling 
the GEANT calibrated models of all the different IBIS sub-systems 
constituting the QM. This model has been used to plan the QM 
calibration, and will be checked against the results of these 
calibration which have occured on September 2000. Figure 1 shows a 
cut-view of this QM model in the X-Z plane. We could see ISGRI in 
green, PiCsIT in black, and the Veto module in red. In blue is the 
IBIS Detector Unit (DU) structure. We could see in figure 2 a 
drawing of the model where the ISGRI QM module is shown in red. 
 
\subsection{IBIS DU and FM Modelling} 
 
In a second step, every IBIS sub-system will create a calibrated 
FM model of its system. The ISGRI, PiCsIT, Veto, and frame will be 
then  ``integrated" to form the virtual Detector Assembly system, 
which will be checked against the calibration to be made in 
December 2000. Then, we will add to the Detector Assembly model 
all the other IBIS sub-systems in order to create the model of the 
full IBIS FM telescope, and the results of the simulations will be 
compared to the IBIS FM calibration (March-April 2001). Figure 3 
shows the result of a run made with our  ``uncalibrated" IBIS full 
model. It is an ISGRI image resulting from this simulation, where 
the source were at the center of the IBIS field of view.

\begin{figure} 
 
\centerline{\psfig{figure=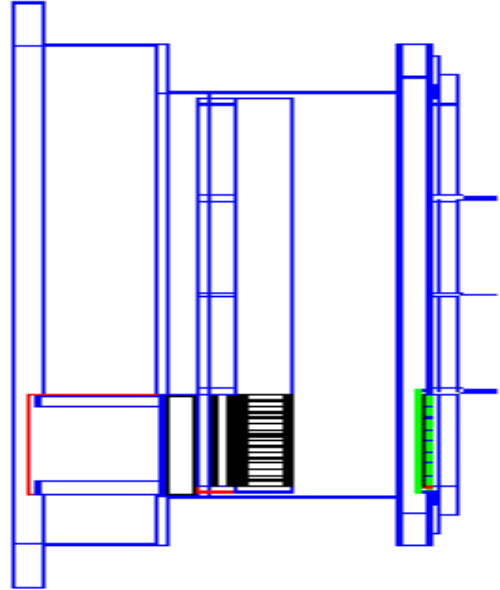,height=85mm,width=85mm}} 
\caption{Side view of the QM GEANT Model .} 
 
\end{figure} 
 
\begin{figure} 
 
\centerline{\psfig{figure=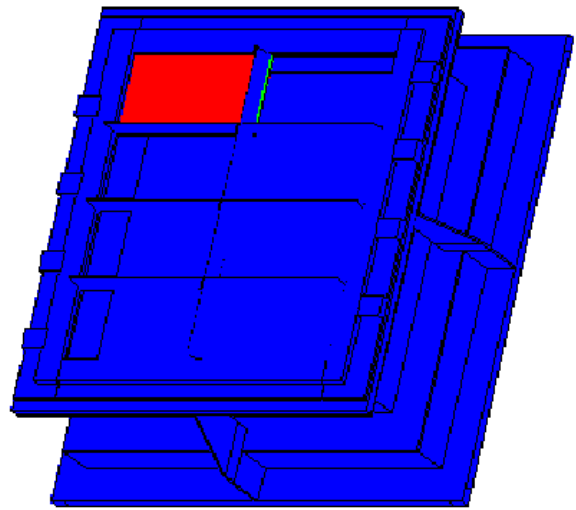,height=80mm,width=85mm}} 
\caption{Top view of the QM GEANT model.} 
 
\end{figure} 
 
\subsection{Response Matrix generation} 
 
A simplified model of  the response matrix will be available for 
testing interfaces with the ISDC system at Fall 2000. The first 
IBIS FM calibrated Mass Model should be available sometimes after 
the IBIS FM calibration, that is at mid 2001. This will enable the 
generation of the first issue of the response matrix, which will 
be checked and refined using the results obtained during the 
PayLoad Ground Calibration done during the ETET period  (mid 2001, 
Carli et al., 2000). A response matrix could be generated for each 
IBIS type of data : ISGRI, PiCsIT single, PiCsIT multiple, Compton 
single, Compton multiple, and Spectral Timing. These matrices will 
be constructed from the simulated IBIS response to an on-axis 
source. The shadowing effects for an off-axis source will be taken 
into account during the image deconvolution process, using 
transparency maps derived for a given off-axis source direction. 
These maps will be checked for some directions using the results 
from the full IBIS FM calibration and the Payload Ground 
Calibration. 
 
\begin{figure} 
\centerline{\psfig{figure=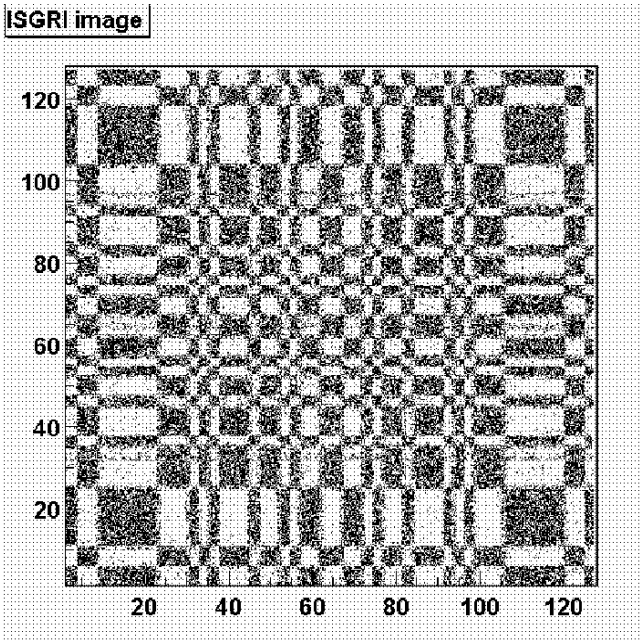,height=85mm,width=85mm}} 
\caption{Simulated ISGRI Image.} 
\end{figure} 
 
\subsection{link with the XSPEC package} 
 
As seen in Figure 4, the results of the IBIS modeling may be 
transformed in XSPEC compatible formats, and used within this 
package. In the figure, we have also taken into account the 
following estimates for the IBIS background values: 
 
ISGRI mode   : 1000 cts/s \newline 
 PiCsIT mode  : 7500 cts/s\newline 
 Compton mode : 100 cts/s 
 
in order to compute the error bars. We have then fit these spectra 
with XSPEC, giving the model spectrum shown in Figure 4. This 
fitted spectrum is composed by a power law with photon index 
$\alpha = -2.4$ and a line centered at an energy of 480 keV with a 
22 keV width, parameters which are consistent with the parameters 
we introduced as input in our Monte-Carlo simulations. 
 
\begin{figure} 
\centerline{\psfig{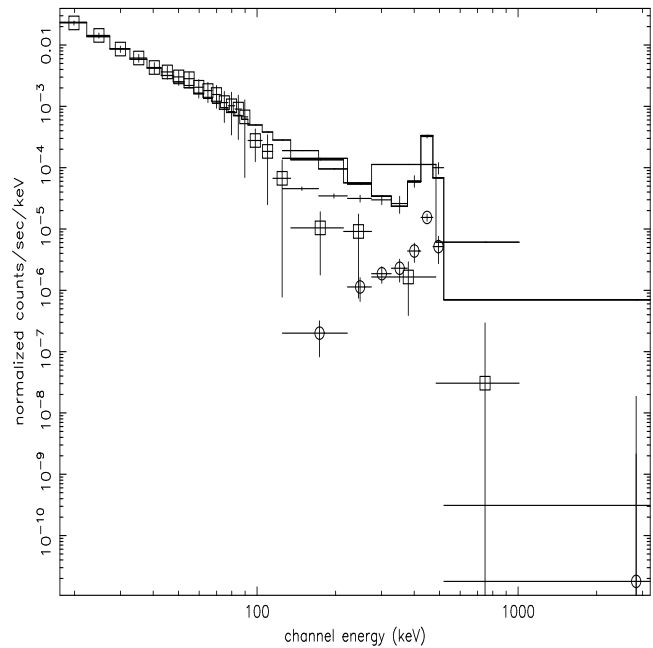}} 
\caption{Simulated IBIS raw spectra as introduced in XSPEC, shown 
with a powerlaw + 511 keV line fit. ISGRI points are indicated by 
squares, PICsIT points by crosses and Compton points by stars. The 
fitted model spectrum composed by a power law with photon index 
-2.4 and a line centered at an energy of 480 keV with a 22 keV 
width is also shown.} 
\end{figure}

\end{document}